\documentclass[submission,copyright,creativecommons]{eptcs}
\providecommand{\event}{MARS 2026} 

\usepackage{graphicx}
\usepackage{xspace}
\hypersetup{hidelinks}
\usepackage{iftex}
\usepackage{xspace}
\usepackage{amsmath}
\usepackage{amssymb}
\usepackage{mathtools, nccmath}
\usepackage{wrapfig}
\usepackage[table,x11names]{xcolor}
\usepackage[labelformat=simple]{subcaption}

\usepackage{tabularx}
\usepackage{booktabs}
\usepackage{multirow}
\usepackage{newtxmath}

\usepackage{siunitx}

\usepackage{uppaal}

\usepackage[
  colorinlistoftodos, 
  textwidth=\marginparwidth, 
  textsize=scriptsize, 
  ]{todonotes}

\usepackage[noabbrev,capitalise]{cleveref}

\title{Safe and Near-Optimal Gate Control:\\ A Case Study from the Danish West Coast\thanks{This research was partly supported by the Interreg North Sea project STORM\_SAFE under Journal ID 41-2-29-23, the Independent Research Fund Denmark under reference number 10.46540/3120-00041B, and the Villum Investigator Grant S4OS under reference number 37819.}}
\author{Martin Kristjansen \quad Kim Guldstrand Larsen \quad Marius Miku\v{c}ionis \quad Christian Schilling
\institute{Department of Computer Science, Aalborg University\\ Aalborg, Denmark}
\email{\{mk,kgl,marius,christianms\}@cs.aau.dk}
}
\def\titlerunning{Safe and Near-Optimal Gate Control}
\def\authorrunning{M. Kristjansen et al.}

\begin{document}
\newcommand{\waterLevel}{\ensuremath{h_w}\xspace}
\newcommand{\oceanHeight}{\ensuremath{h_s}\xspace}
\newcommand{\fjordHeight}{\ensuremath{h_f}\xspace}
\newcommand{\dd}{\mathrm{d}}
\newcommand{\bottomHeight}{\ensuremath{h_b}\xspace}
\newcommand{\gateWidth}{\ensuremath{w_g}\xspace}
\newcommand{\gateHeight}{\ensuremath{h_g}\xspace}
\newcommand{\crossSingle}{\ensuremath{A_g}\xspace}
\newcommand{\crossAll}{\ensuremath{\crossSingle^\mathit{all}}\xspace}
\newcommand{\oceanFlow}{\ensuremath{Q_\mathit{gates}}\xspace}
\newcommand{\streamInflow}{\ensuremath{Q_\mathit{streams}}\xspace}
\newcommand{\areaFjord}{\ensuremath{A_f}\xspace}
\newcommand{\strategy}{\ensuremath{\sigma}\xspace}
\newcommand{\horizon}{\ensuremath{{k}}\xspace}
\newcommand{\period}{\ensuremath{{p}}\xspace}

\newcommand{\state}{\ensuremath{s}\xspace}
\newcommand{\states}{\ensuremath{\mathbf{S}}\xspace}
\newcommand{\action}{\ensuremath{a}\xspace}
\newcommand{\actions}{\ensuremath{\mathbf{A}}\xspace}
\newcommand{\reward}{\ensuremath{r}\xspace}
\newcommand{\reals}{\ensuremath{\mathbb{R}}\xspace}

\maketitle

\begin{abstract}
Ringk\o{}bing Fjord is an inland water basin on the Danish west coast separated from the North Sea by a set of gates used to control the amount of water entering and leaving the fjord.
Currently, human operators decide when and how many gates to open or close for controlling the fjord's water level, with the goal to satisfy a range of conflicting safety and performance requirements such as keeping the water level in a target range, allowing maritime traffic, and enabling fish migration.
In this paper, we present a digital twin of the fjord's water level, modeled in the tool \uppaal \stratego.
We then use this digital twin along with forecasts of the sea level and the wind speed to learn a gate controller in an online fashion.
We evaluate the learned controllers under different sea-level scenarios, representing normal tidal behavior, high waters, and low waters.
Our evaluation demonstrates that, unlike a baseline controller, the learned controllers satisfy the safety requirements, while performing similarly regarding the other requirements.
\end{abstract}

\section{Introduction}

\begin{figure}[ht!]
    \centering
    \includegraphics[width=0.95\textwidth]{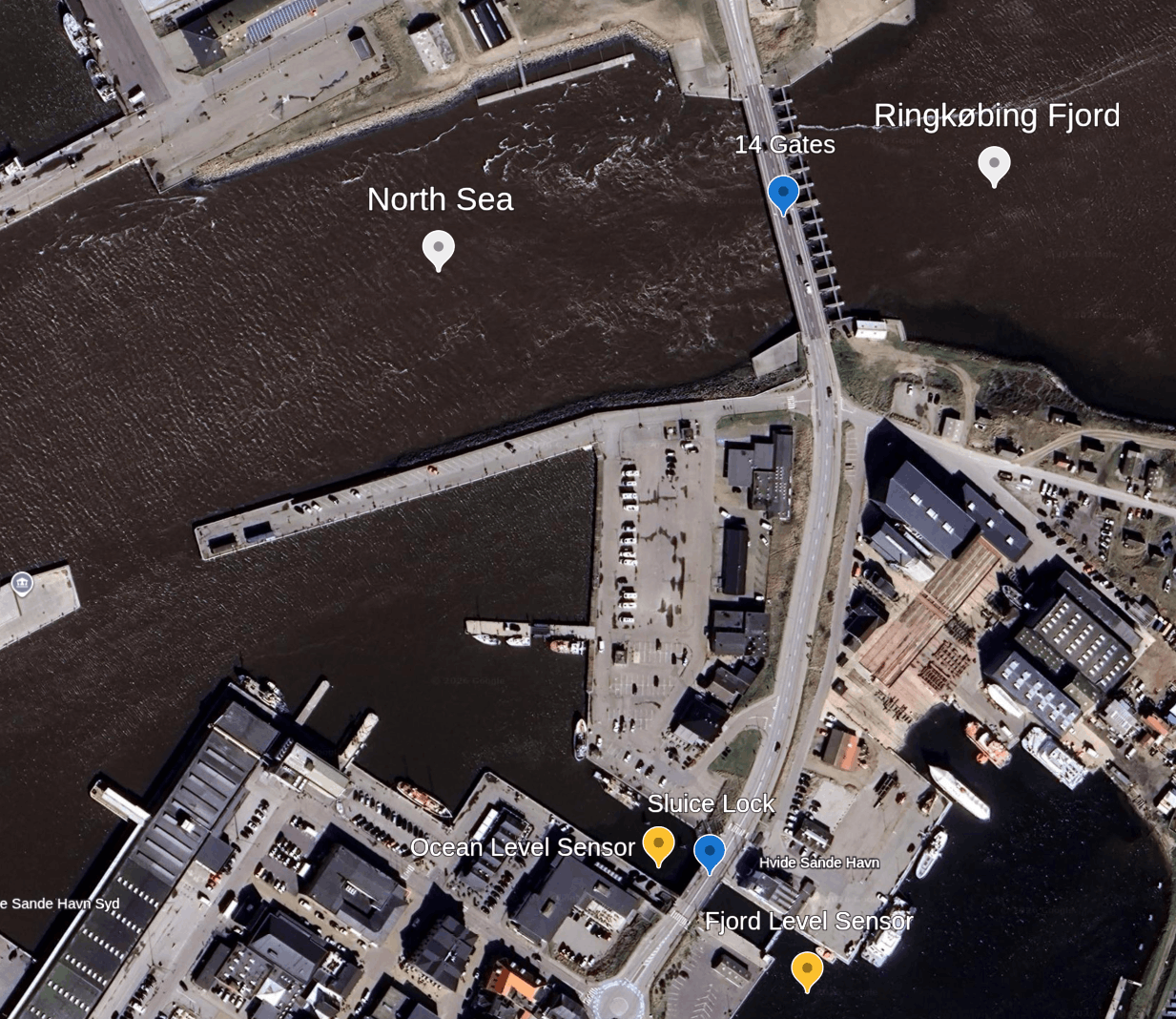}
    \caption{Satellite image of the 14 gates and the sluice lock at Hvide Sande harbor.}
    \label{fig:sluices}
\end{figure}

\begin{figure}[ht!]
    \centering
    \includegraphics[width=0.95\textwidth]{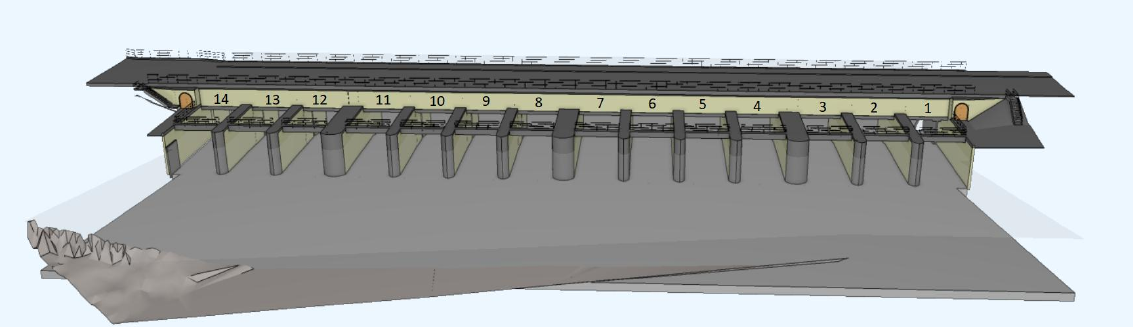}
    \caption{A 3D model provided by the DCA of the 14 sluices.}
    \label{fig:sluices2}
\end{figure}

Ringk\o{}bing Fjord is an inland water basin (henceforth called fjord) on the Danish west coast with a surface area of \qty{290}{km^2} and a depth of up to \qty{5}{m}.
As visualized in \cref{fig:sluices}, the fjord is connected to the North Sea through a set of 14 individually controllable gates in a water dam shown in \cref{fig:sluices2} and a separate but close-by sluice lock for maritime traffic, situated in the town of Hvide Sande.
Two sensors are placed on each side of the sluice lock to measure the water levels of the fjord and the sea.
The operation of gates serves multiple objectives.
Most importantly, the water level of the fjord must be within a safe range to avoid damage to the wildlife (e.g., by drying out nesting sites) or property (through flooding).
Fish must be able to migrate between the sea and the fjord, and incoming maritime traffic must have easy access to the sluice lock, i.e., avoid interfering current from the gates.

This paper is motivated by the need of modern control support for this critical infrastructure, as the gates are currently operated manually according to a set of operation principles, where the operators have room to decide when to open and close the gates.
Moreover, the operators face the challenge of rising sea levels as well as an increase in the amount of water coming from streams and the surrounding catchment area because of more frequent and heavy rainfalls due to climate change.
We first present a model of how the fjord's water level is affected by the sea's current water level and the number of open gates.
Then we use this model as a digital twin (i.e., simulation model) in reinforcement learning within the tool \uppaal \stratego~\cite{stratego}, where we learn a controller for operating the gates so that a range of safety and performance requirements is respected.
We consider the weather forecast when learning a controller, but a forecast is only valid (or has a small uncertainty) for the near future.
Therefore, we use an online approach, where the controller is updated whenever a new weather forecast is available.
Finally, we compare the learned controllers with a baseline controller on multiple performance requirements.

\paragraph{Problem Description.}
Our focus in this paper is on the operation of the gates, whose main objective is to ensure a balanced water level of the fjord.
High water levels cause damage to the local residential areas, and low water levels negatively impact the wildlife.
Besides a general protection against storm water from the sea, the gates are used to control when water is let out to the sea if the fjord's water levels are too high (typically due to precipitation and inflow from the mainland) and when water is let in from the sea (typically after periods with little precipitation).
One challenge is that the sea level changes with the tides, and the generally rising sea levels reduce the time window when water can be let out to the sea.

\paragraph{Requirements.}

\begin{table}[ht!]
	\caption{Requirements for the operation of gates.}
	\label{tab:requirements}
	\centering
	\begin{tabular}{rll}
      \toprule
      \textbf{Criticality} & \textbf{Id} & \textbf{Requirement} \\
      \midrule
      \multirow{3}{*}{Safety} &
      S.1 & The fjord water level must be between \qty{0.00}{m} and \qty{0.25}{m} DVR90 \\
      & S.2 & Allow water into the fjord only when the wind speed is above \qty{8}{m/s} \\
      & S.3 & Gates are closed when the water level difference is above \qty{1}{m} \\
      \midrule
      \multirow{3}{*}{Performance} &
      P.1 & Enable fish migration \\
      & P.2 & Minimize the number of operations (wear and tear)\\
      & P.3 & Minimize the waiting time for boat passage\\
      \bottomrule
	\end{tabular}
\end{table}

The Danish Coastal Authority (DCA)\footnote{\url{https://kyst.dk/}} is responsible for the operation of this infrastructure.
From the informal specification given by the DCA, we derived three safety requirements and three performance requirements, all summarized in \cref{tab:requirements}.

The first safety requirement~S.1 establishes the acceptable water-level range relative to DVR90 (Danish Vertical Reference defined in 1990).
Thus, the fjord's water level must be inside a safety range of \qty{0.0}{m} and \qty{0.25}{m} with respect to DVR90.
The bottom of the gates is at \qty{-4.1}{m} (below DVR90).
Regarding safety requirement~S.2, a wind speed of \qty{8}{m/s} is needed to guarantee that the water is properly mixed when letting water from the sea into the fjord.
(However, even with wind speeds below \qty{8}{m/s}, a single gate may still be open to allow for fish migration.)
Finally, the safety requirement~S.3 ensures that the gates are not used when the water levels differ by \qty{1}{m} or more.
When the difference is that great, the gates might not be able to withstand the forces, thus damaging the gates permanently.

Apart from the safety requirements, we must consider the performance requirements for a good operation of the gates.
Performance requirement~P.1 says that at least one gate should be open as long as possible for fish to migrate between the fjord and the sea when the water level difference between the sea and the fjord is within $\pm\qty{0.1}{m}$.
This applies both during the day and at night.
There is no equipment installed that measures whether fish are swimming in and out of the fjord, so the DCA uses the time gates are opened as a proxy for evaluating fish migration.
Performance requirement~P.2 expresses that the gates should be opened and closed as few times as possible, as this wears the hardware.
Finally, the performance requirement~P.3 aims to reduce currents and allow boats to navigate safely to the sluice lock; therefore, the gates are to be closed for incoming traffic.

Improving some of these performance requirements may negatively affect others, e.g., closing the gates for incoming boats reduces the time for fish migration.
Any controller for the gates must fulfill all safety requirements before considering the three performance requirements.
However, different controllers can fulfill the safety requirements but handle the performance requirements differently.

\subsection{Related Work}

\textbf{Modeling of Ringk\o{}king Fjord.}
Ringk\o{}king Fjord has been studied by multiple previous works.
Nielsen et al.~\cite{nielsen2005simple} proposed a model to predict the effect on the fjord's water level and the salinity given the sea level and the wind; the model was created using the Manning method~\cite{manning}.
The method produced good results for the water level when the wind was not strong, while the predicted salinity sometimes deviated significantly from the true value.
In this paper, we do not explicitly take salinity into account, in part because the available data is only provided once per week, which we deem too infrequent for our online learning approach.
Later work investigated the optimal amount of salinity~\cite{bryhn2009multi}.
This included the nutrients and the salinity, as well as the effect of sudden changes in salinity.
Possible options to stabilize the salinity were suggested, such as adding a saltwater pumping station.
Another related study used a coupling of a so-called MIKE SHE hydrological model and HIRHAM regional climate model to compute the effect on the catchment area around Skjern lake, which leads to Ringk\o{}bing Fjord~\cite{larsen2014results}.
We do not currently take a sophisticated model of the water inflow from the mainland into account.

\textbf{Online control.}
Our online approach to control the gates shares similarities with model predictive control (MPC)~\cite{borrelli2017predictive,rawlings2017model}, where a model of the system is given up-front.
At operation time, a good control strategy for the current state is derived from the model, and then only the first control action is executed, leading to the next system state, from where the process repeats.
The main difference in our case is that we repeatedly learn a controller online, which is infeasible in MPC because it needs to react quickly and thus uses fast control heuristics.
Once available, the learned controller is thus expected to yield better performance.
Our approach is particularly suited for slow-moving systems (e.g., in this case, there are ten minutes between any two control actions).

\paragraph{Outline.}
The paper is organized as follows: Section~2 presents the concept of online reinforcement learning. 
Section~3 presents the model for the fjord's water level and how it is affected by the sea's water level and the number of open gates.
Section~4 presents the \uppaal \stratego model of the system.
Section~5 presents the experimental setup and Section~6 presents the results.
Section~7 concludes the paper and lists directions for future work.

\section{Online Reinforcement Learning}

\begin{figure}[ht!]
    \centering
    \includegraphics[width=0.5\textwidth]{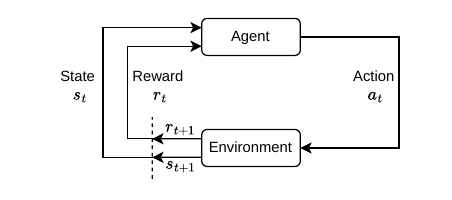}
    \caption{General setup of reinforcement learning.}
    \label{fig:reinforcement}
\end{figure}

To obtain a controller for operating the 14 gates, we use reinforcement learning~\cite{reinforcement}, which consists of interactions between an environment and an agent, as shown in \cref{fig:reinforcement}.
These interactions take place in episodic runs consisting of multiple steps.
In each step, being in some system state~$\state_t\in\states$, the agent chooses an action~$\action_t\in\actions$, followed by the environment leading to the next state~$\state_{t+1}$ and then providing a reward~$\reward_{t+1}\in\reals$ (or cost) to the agent.
Thus, a run of the system from some initial state~$\state_0$ has the form
\begin{equation*}
    \state_0,\action_0,\reward_1,\state_1,\action_1,\reward_2,\state_2,\action_2,\ldots
\end{equation*}

The aim of the agent is to maximize the accumulated reward over each run.
To achieve that, the agent must learn which action to perform given the current state of the system.
Specifically, we want the agent to learn a deterministic and memoryless strategy \strategy, which is a function of the form $\strategy \colon \states \rightarrow \actions$.

We face three major challenges in this use case.
First, operation with the physical environment for reinforcement learning is not feasible, as the consequences of bad operations are learned by doing those operations on the real system with actual consequences on the environment.
Second, if we were to learn a strategy that works this year, it may not work in twenty years due to rising sea levels.
Third, using the weather forecast as input to the controller would make the domain infeasibly large, as the controller must optimize for all possible forecasts.
To accommodate for these challenges, we create a \emph{digital twin} to simulate the environment (e.g., the water levels of the sea and the fjord) and use \emph{online reinforcement learning} to periodically update the strategy during operation.
In this work, we use Q-learning with partition refinement, in which the latter discretizes the continuous state space during learning, enabling Q-learning to be applied to continuous variables~\cite{play_ball}.
The online learning ensures that the domain shift (rising sea levels) is accounted for automatically, and it allows us to encode the weather forecast implicitly into the environment instead of exposing it to the controller explicitly.

In online reinforcement learning~\cite{stompc}, first an initial strategy is learned as usual to maximize the accumulated reward up until some horizon~\horizon.
Then, at operation time, after controlling the system for a time period $\period \leq \horizon$ and while monitoring it, the digital twin is updated based on the observed behavior, which is then used to compute a new strategy.
This process then repeats.
That approach has proved successful in multiple domains, ranging from fleet management of robots~\cite{fleet_management}, where the aim is to minimize the time the robots need to complete a set of tasks, to operating water detention ponds~\cite{water_pond}, where the aim is to control the water level of ponds using the weather forecast.

In this work, we feed the weather forecast to the digital twin, and the horizon of the forecast defines how far in the future the digital twin is able to estimate the system's state.
The weather forecast is valid for three days, which sets the upper bound on the horizon~$k$ during learning. 
We use three days as the horizon, as it enables the system to learn preemptive actions for all known upcoming water levels.
A new and updated forecast is provided every six hours, yielding a time period~$p$ of at least six hours.
Once a new forecast arrives, we start learning a new strategy to replace the current strategy.
However, we continue operating with the current strategy until the new strategy is available.
Moreover, the digital twin must estimate the water level in the fjord.
In the next section, we present our model for this task.

\section{Model for Predicting Water Levels}
\label{sec:water_model}
The amount of water in the fjord is determined by how much water flows between the sea and the fjord as well as how much water enters the fjord from surrounding streams.
We denote the flow at the gates separating the fjord and the sea as \oceanFlow and the flow from the streams as \streamInflow.
The unit of flow is \unit{m^3/s}, indicating the change in volume relative to time.
The total flow is
\begin{equation*}
	Q_\mathit{total} = \oceanFlow + \streamInflow.
\end{equation*}
We say that water is entering (resp.\ leaving) the fjord if $Q_\mathit{total}$ is positive (resp.\ negative).
Next, we present the models we use to estimate the flows \oceanFlow and \streamInflow.

For simplicity, we assume that the water levels of the sea and the fjord can be described by a single value.
(In reality, this is not the case even for the fjord, as the wind is strong enough to cause different water levels depending on which end of the fjord is measured.)
This assumption of the fjord having a flat surface also implies that, when water flows through the gate, the water level in the fjord changes uniformly.
We denote the sea level using~\oceanHeight and the level of the fjord using~\fjordHeight (in meters) with respect to the reference height DVR90.
For the sea level, we rely solely on weather forecasts when modeling that level, as it is governed by tides and storms.

The flow between the sea and the fjord is determined by two factors: the difference in water levels between the two bodies of water and the cross-section of the opening between the bodies of water (i.e., the total area in which water can flow through the opened gates).
We approximate the relationship between the water flow, height difference, and the cross-section of the opening with the equation
\begin{equation*}
    \oceanFlow = K \cdot A \cdot \sqrt{|\Delta h|},
\end{equation*}
where $K$ is a flow constant, $A$ is the area of the cross-section that the water flows through, and $\Delta h$ is the height difference of the water levels.
This equation is internally used by the DCA as a coarse estimate of the water flow.
If \oceanFlow is positive, water flows from the sea to the fjord, while if negative, water flows from the fjord to the sea.
The flow constant $K$ depends on the direction of the flow: the coefficients to be $K_\mathit{inflow} = 3.8 \unit{m^{1/2}/s}$ when flowing from the sea to the fjord and $K_\mathit{outflow} = -3.5 \unit{m^{1/2}/s}$ when flowing from the fjord to the sea.
The intuition behind $K_\mathit{inflow}$ having a greater absolute value is that when water enters the fjord, there is a greater mass pushing water inwards.
When water leaves the fjord, there is less mass to push the water out, resulting in a lower absolute value for the flow constant.

\begin{figure}[ht!]
    \centering
    \includegraphics[height=55mm,keepaspectratio]{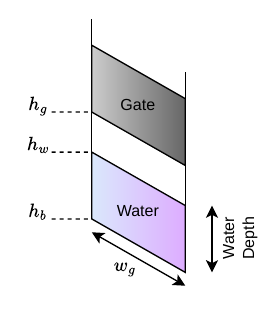}
    \caption{View of a single open gate and the values used to compute the area through which water flows between the fjord and the sea.}
    \label{fig:openGate}
\end{figure}

To estimate the cross-sectional area of the opening that the water passes when moving between the sea and the fjord, we must know the water levels, the number of open gates, the width of those gates (because they are rectangular), and how much these gates are opened.
The gates touch the bottom of the fjord when closed and move up vertically when opened.
Considering a single gate, we denote its bottom with respect to DVR90 as \bottomHeight and its width as \gateWidth, as illustrated in \cref{fig:openGate}.
Then, the height of the cross-section is either given by the water level \waterLevel or the bottom of the gate \gateHeight, whichever is lower.
There must be a height difference between the sea and the fjord for a flow to occur.
Consequently, we use the greatest water level of the two bodies of water as the water level value $\waterLevel = \max(\fjordHeight, \oceanHeight)$, representing the side from which water is flowing.
Thus, the cross-section of a single gate \crossSingle is
\begin{equation*}
	\crossSingle = \gateWidth \cdot (\min(\gateHeight, \waterLevel) - \bottomHeight),
\end{equation*}
and we denote the sum of all cross-section areas of all gates as~\crossAll.

The second contributing flow \streamInflow is modeled using historical data for approximating the average flow in a given calendar month.
Thus, the inflow of fresh water in January is different from the value in August.
This approximation is based on empirical measurements\footnote{\url{https://hip.dataforsyningen.dk/} was used to obtain the flow of all streams flowing into the fjord.}, but such an approximation does not consider situations like extreme rainfall or the melting of deep layers of snow, which then fills up the fjord quickly.
Nonetheless, it is a useful approximation for operation under normal conditions.

Having a model for $Q_\mathit{total}$, we can estimate the change in the fjord's water level.
Exploiting the assumption that the fjord's surface area~\areaFjord is flat, and further assuming that~\areaFjord is constant for different water levels, the height of the fjord's water level~\fjordHeight at a given time point $t$ can be approximated as
\begin{equation*}
	\fjordHeight(t) = \frac{V(t)}{\areaFjord},
\end{equation*}
where $V(t)$ is the volume of the water in the fjord at time~$t$.
Then, the change in the fjord's water level with respect to time $t$ is

\begin{equation*}
	\frac{{\dd \fjordHeight(t)}}{\dd t} = \frac{\dd V(t)}{\dd t}\cdot\frac{1}{\areaFjord } = Q(t) \cdot\frac{1}{\areaFjord},
\end{equation*}
where $Q(t)$ is the flow at time $t$.
Since we have \oceanFlow and \streamInflow, we can further specify how the fjord's water level is affected, such that the state of the gates is taken into consideration.
The change in the fjord's water level is given as

\begin{equation}\label{eq:fjordLevel}
	\frac{\dd \fjordHeight(t)}{\dd t} = \begin{cases} 
		\dfrac{K_\mathit{inflow} \cdot \crossAll(t) \cdot \sqrt{\oceanHeight(t) - \fjordHeight(t)}}{\areaFjord} + \dfrac{\streamInflow(t)}{\areaFjord}   &\text{ when  } \oceanHeight \geq \fjordHeight    \\[2ex]
		\dfrac{K_\mathit{outflow} \cdot \crossAll(t) \cdot \sqrt{\fjordHeight(t) - \oceanHeight(t)}}{\areaFjord} + \dfrac{\streamInflow(t)}{\areaFjord} & \text{ otherwise}.
	\end{cases}
\end{equation}

\section{Modeling Using \uppaal \stratego}\label{sec:modeling}

We use the tool \uppaal \stratego~\cite{stratego} to describe our digital twin.
Specifically, we model the system as a network of five \emph{stochastic hybrid automata}~\cite{smc}.
One automaton models the water level of the fjord, another automaton models the controller, and the third automaton models the interaction with boats waiting to pass.
The remaining automata are responsible for surveilling the requirements from Table~\ref{tab:requirements}.
Once modeled, \uppaal \stratego can apply reinforcement learning to obtain near-optimal strategies using a combination of Q-learning~\cite{reinforcement} and partition refinement~\cite{play_ball}.
The latter discretizes the continuous state space during the learning process, enabling the application of Q-learning for continuous variables such as the water level.
The output of \uppaal \stratego is a decision tree, which maps any system state to the action with the highest Q-value for that state.

\begin{figure}[ht!]
	\centering
	\includegraphics[width=0.95\textwidth]{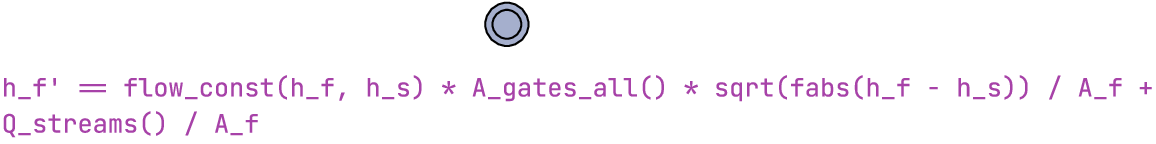}
	\caption{Automaton for the fjord's water level~\uppClock{h_f}. It makes use of the built-in function \uppFunc{fabs}, returning the absolute value of a floating-point expression.}
	\label{fig:waterlevel}
\end{figure}

\paragraph{Automaton model.}
In our \uppaal \stratego model, one time unit corresponds to one minute.
We define a global hybrid clock \uppClock{h_f} (i.e., a continuous variable), which reflects the current water level~$h_f$ of the fjord.
To model the evolution of \uppVar{h_f}, we use the automaton \uppProc{FjordWaterLevel} shown in \cref{fig:waterlevel} consisting of a single location.
As can be seen, \uppaal \stratego allows us to directly encode the (first-order) ordinary differential equation (ODE) as presented in \cref{eq:fjordLevel}, which \uppaal \stratego solves by numerical integration.
To formulate the rate, we make use of C-like helper functions to determine the flow constant, the cross-section at the gates, and the flow from incoming streams; using this function interface allows for a concise visual understanding of the model while still being able to conveniently extend the functionality later.

The inflow of streams~\uppFunc{Q_streams()} is, as mentioned earlier, an average flow of the incoming streams for any given month.
This flow is calculated by using a base value \qty{3558}{m^3/min} multiplied by a monthly weight shown in \cref{tab:weight}.
Thus, the flow in January is estimated $\qty{3558}{m^3/min} \cdot 1.45 = \qty{5159.10}{m^3/min}$.

\begin{table}[ht!]
	\caption{The monthly weights of the water inflow from the fjord's surrounding streams.}
	\label{tab:weight}
	\centering
	\begin{tabular}{l|llllllllllll}
      \toprule
      \textbf{Month} & \textbf{Jan} & \textbf{Feb} & \textbf{Mar}  & \textbf{Apr}  & \textbf{May}  & \textbf{Jun}  & \textbf{Jul}  & \textbf{Aug}  & \textbf{Sep}  & \textbf{Oct}  & \textbf{Nov}  & \textbf{Dec}  \\
      \midrule
      \textbf{Weight} & 1.45 & 1.39  & 1.30  & 1.01  & 0.82  & 0.71  & 0.66  & 0.64  & 0.72  & 0.87  & 1.12  & 1.29 \\
      \bottomrule
	\end{tabular}
\end{table}

To utilize historical data, such as weather forecasts and wind speeds, we have written a shared C-library, which we import into our \uppaal \stratego model as shown in \cref{lst:import}.
The first imported function \uppFunc{get_config()} reads a configuration file containing the time and date from which to use the historical data.
Thus, the same \uppaal \stratego model can be used for different time periods as the dates used are determined by an external configuration file.
The next three functions are used to obtain historical data about the water levels or the wind speed as a function of time.
(Data on the fjord's water level is only used to initialize the model.)
The remaining imported functions are used to read the values imported through the configuration.
\begin{uppaalcode}[caption={Import of external functions},label={lst:import}]
import "./libcsvReaderLib.so" {
    int get_config();
	int get_ocean_level(int year, int month, int day, int hour, int minute);
	int get_fjord_level(int year, int month, int day, int hour, int minute);
	double get_wind_speed(int year, int month, int day, int hour, int minute);
	int get_year();
	int get_month();
	int get_day();
	int get_hour();
	int get_minutes();
};
\end{uppaalcode}

\medskip

\begin{figure}[ht!]
    \centering
    \includegraphics[width=\textwidth]{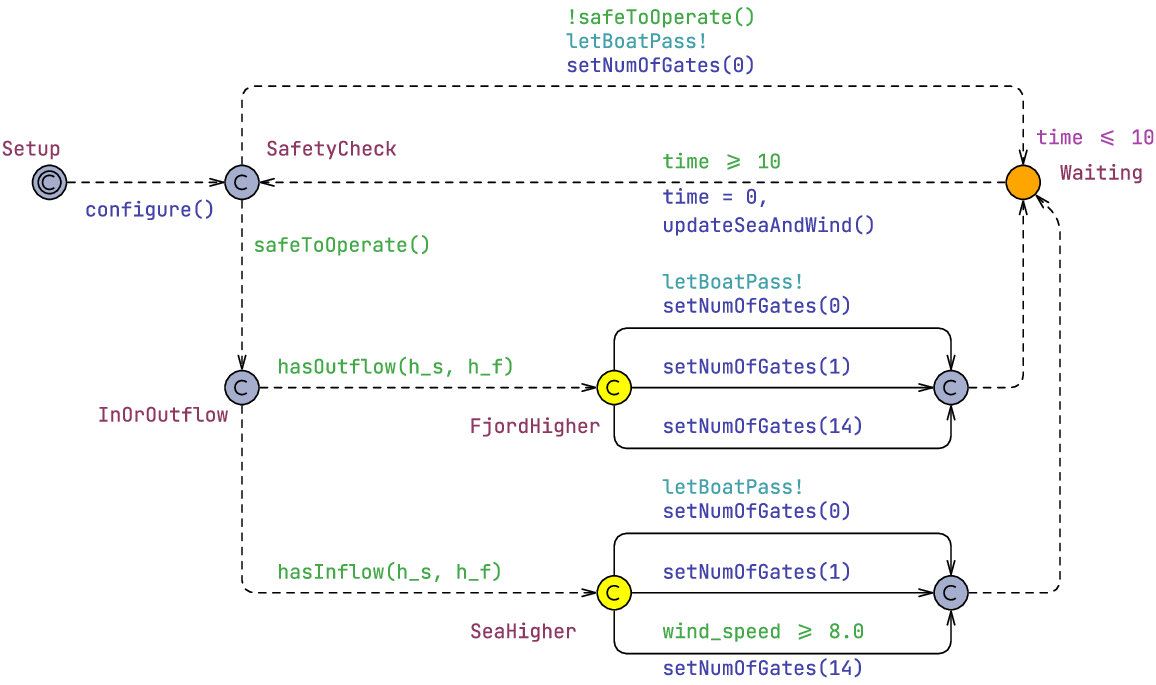}
    \caption{Automaton for the gate controller. The dashed lines represent uncontrollable (environment) decisions, whereas the solid lines represent actions by the controller. Guards are written in {\color{uppGuardColor} green}, updates in {\color{uppUpdateColor} blue}, invariants in {\color{uppInvColor} magenta}, and synchronization channels in {\color{uppSyncColor} cyan}.}
    \label{fig:learner}
\end{figure}

The second automaton in our model is the \uppProc{Controller} automaton, which is shown in \cref{fig:learner}.
This automaton periodically controls the number of gates to open or close.
Any decision is based on sensor readings of the water levels and wind speed, which are updated every 10 minutes.
We design the controller to choose an action (possibly a non-action) whenever new sensor readings are provided.

In the \uppProc{Controller} automaton, we do not allow time to pass in \emph{committed} locations (denoted with a \uppVar{C}).
Moreover, when any automaton is in a committed location, the next edge taken must be from a committed location.
When leaving the initial location (double circle) \uppLoc{Setup}, the function \uppFunc{configure()} is evaluated, which calls several of the imported functions from \cref{lst:import} such as \uppFunc{get_config()}.
The initial value of the water levels and the date and time are imported, including the \uppVar{month} variable used to determine the flow from streams.
In the location \uppProc{SafetyCheck}, the transition taken depends on whether it is safe to open the gates.
According to requirement~S.3, the gates must be closed if the difference in water level is above \qty{1}{m}, which is determined by the guard function \uppGuard{safeToOperate()}.
If it is safe to operate, we use the next location to determine whether there is an inflow (sea to fjord) or an outflow (fjord to sea).
No matter which way the water flows, in the yellow locations, the controller selects between having no gates, a single gate, or all 14 gates open.
In this model, we limited the controller to those three actions.
The rationale is that these are the most commonly used modes in the current practice.

If the option of no open gates is chosen, a broadcast message \uppSync{letBoatPass} is sent, as this gate configuration allows any waiting boats to reach the sluice lock.
One aspect to note is that to open all 14 gates in the (yellow) \uppLoc{SeaHigher} location, the wind speed must be \qty{8}{m/s} or more, respecting the safety requirement~S.2.
When the number of open gates has been chosen, the automaton stays in the (orange) \uppLoc{Waiting} location for 10 minutes.
Finally, the new water levels as well as the wind speed are read via the function \mbox{\uppFunc{updateSeaAndWind()}} and the next control period starts.

Our controller has a relatively small action space (three actions). However, the complexity stems from the total amount of actions over the course of a scenario (every ten minutes for three days), leading to a total of $3^{432}$ possible action vectors.

\begin{figure}[ht!]
	\centering
	\begin{minipage}[t]{.3\textwidth}
		\centering
		\includegraphics[width=\textwidth]{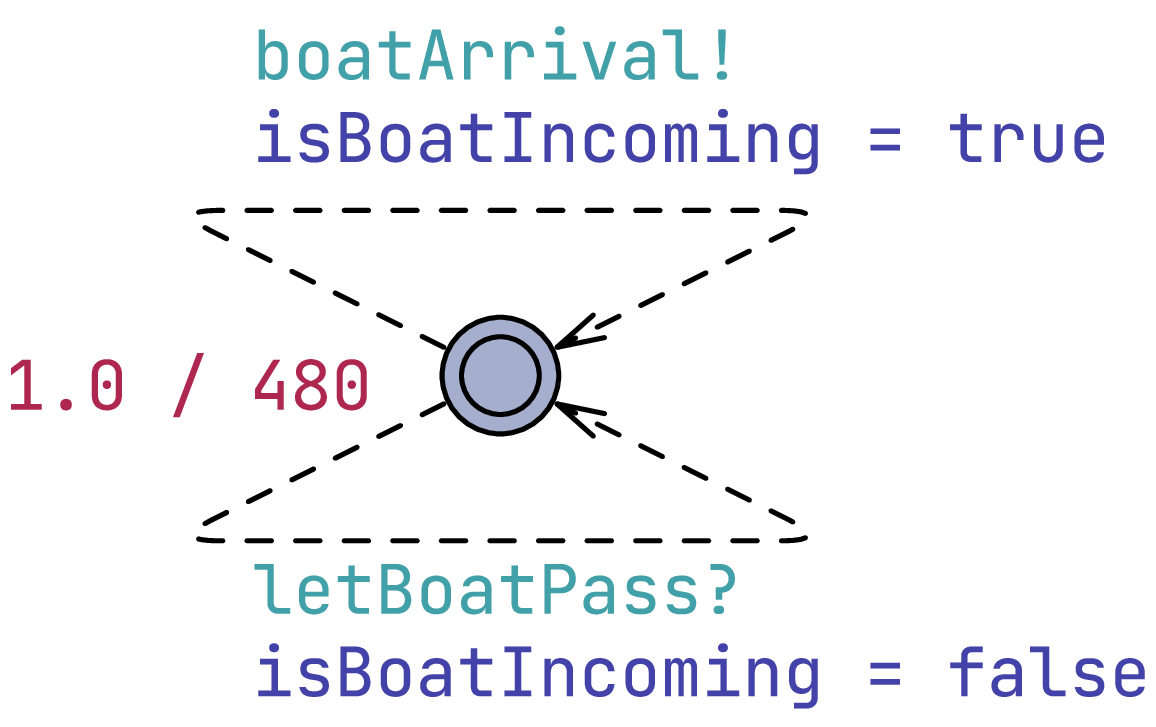}
		\caption{Automaton modeling boat arrivals.}
		\label{fig:boats}
	\end{minipage}\hfill%
	\begin{minipage}[t]{.68\textwidth}
		\centering
		\includegraphics[width=\textwidth]{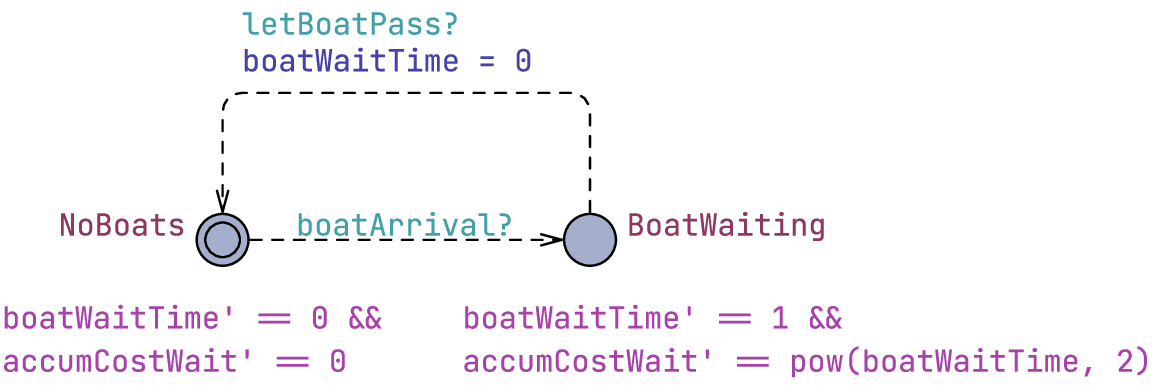}
		\caption{Automaton measuring the time a boat has waited for passage and the accumulated squared waiting time of all boats.}
		\label{fig:boatsWaiting}
	\end{minipage}
\end{figure}

The third automaton, shown in \cref{fig:boats}, models the arrival and leaving of boats.
This automaton consists of one location and two uncontrollable edges: one signaling that a boat arrives and another one to notify that the boat is allowed passage.
For the sake of simplicity, we model the stochastic boat arrival using an exponential distribution with the rate~$1/480$, and we signal the arrival using the broadcast channel \uppSync{boatArrival}.
Thus, the average time between arrivals is 480 minutes (eight hours), which is realistic according to the current gate operators (and it can be replaced by an even more realistic distribution in the future).
Whenever the gates are closed, arriving boats do not have to wait; this is modeled with the second edge, which receives a signal via \uppSync{letBoatPass}.

The remaining two automata measure properties of the model.
One automaton, shown in \cref{fig:boatsWaiting}, manages the rates of two hybrid clocks (essentially stop watches): \uppClock{boatWaitTime}, a measuring how long the current boat has waited, and \uppClock{accumCostWait}, measuring the accumulated squared waiting times of all boats.
The latter is motivated because we assume an exponential loss in patience when the boat has to wait for too long.
\uppClock{boatWaitTime} is reset to zero every time a boat can reach the sluice lock.

\begin{figure}[ht!]
	\centering
	\includegraphics[width=0.32\textwidth]{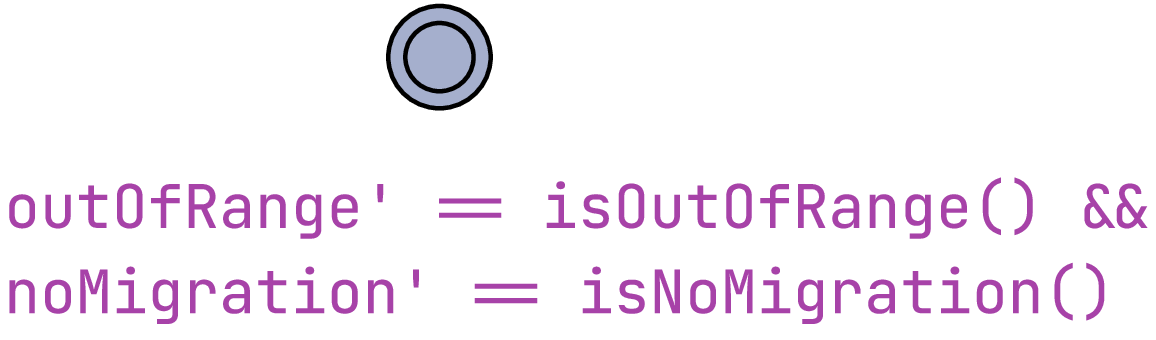}
	\caption{Automaton measuring the durations of (1) the fjord level being unsafe and (2) fish migration being prohibited due to the gates being closed but otherwise being allowed by the water levels.}
	\label{fig:obs}
\end{figure}

The final automaton, shown in \cref{fig:obs}, uses two stop watches to measure the time the water level of the fjord is outside its safe range and the time fish are not allowed to migrate.
The rates of these stopwatches are obtained by evaluating two Boolean functions.
The clock \uppClock{noMigration} has a rate of 1 when all gates are closed while the difference in water levels is within $\pm\qty{0.1}{m}$.
This represents the case in which fish would have optimal conditions for migration, but the gates are closed for some reason.

\paragraph{Learning objective.}
Given the network of automata described above, we define the learning query for managing the gates.
To learn a near-optimal controller, we need a reward function.
In our case, it is more convenient to define a cost function and try to minimize it.
The cost of a state is determined by four parameters: 1) the amount of time spent outside the safe range, 2) the squared waiting times of arriving boats, 3) how many times the configuration of the gates have changed, and 4) the amount of time where the gates could have been opened for fish migration.
The third parameter is handled by a global variable \uppClock{C_gates}, which is incremented in the function \uppFunc{setNumOfGates(...)} whenever the gate configuration is changed.
For now, we do not distinguish whether the configuration changes one or all gates.
The other parameters have been explicitly shown in the automata before.
Each of those four parameters is given an individual weight $w_1$--$w_4$, which we use in the following learning query:

\begin{uppaalcode}
	strategy optOperation = 
        minE(w_1*outOfRange + w_2*accumCostWait + w_3*C_gates + w_4*noMigration) 
        [<=THREE_DAYS]
        {Controller.FjordHigher, Controller.SeaHigher, isBoatIncoming} -> 
        {h_f, h_s, wind_speed} : <>  global_time >= THREE_DAYS
\end{uppaalcode}

This query aims to obtain a strategy minimizing the value of the cost function (\uppProc{minE}).
Given that all weights are positive, the used cost function is monotonic.
The controller must determine how many gates should be open in every 10-minute period, based on whether the \uppProc{Controller} automaton is in either the \uppLoc{FjordHigher} or \uppLoc{SeaHigher} location, and the value of \uppVar{isBoatIncoming}.
Then, using partition refinement, the decision can further depend on the water levels of the sea and the fjord along with the wind speed.
The horizon of the query is three days, which ends one training episode.
In the next section, we explain the experimental setup in which we execute this query for learning a strategy.

\begin{figure}[ht!]
  \centering
  \begin{subfigure}{0.7\textwidth}
    \centering
	\includegraphics[width=\textwidth,height=65mm,keepaspectratio]{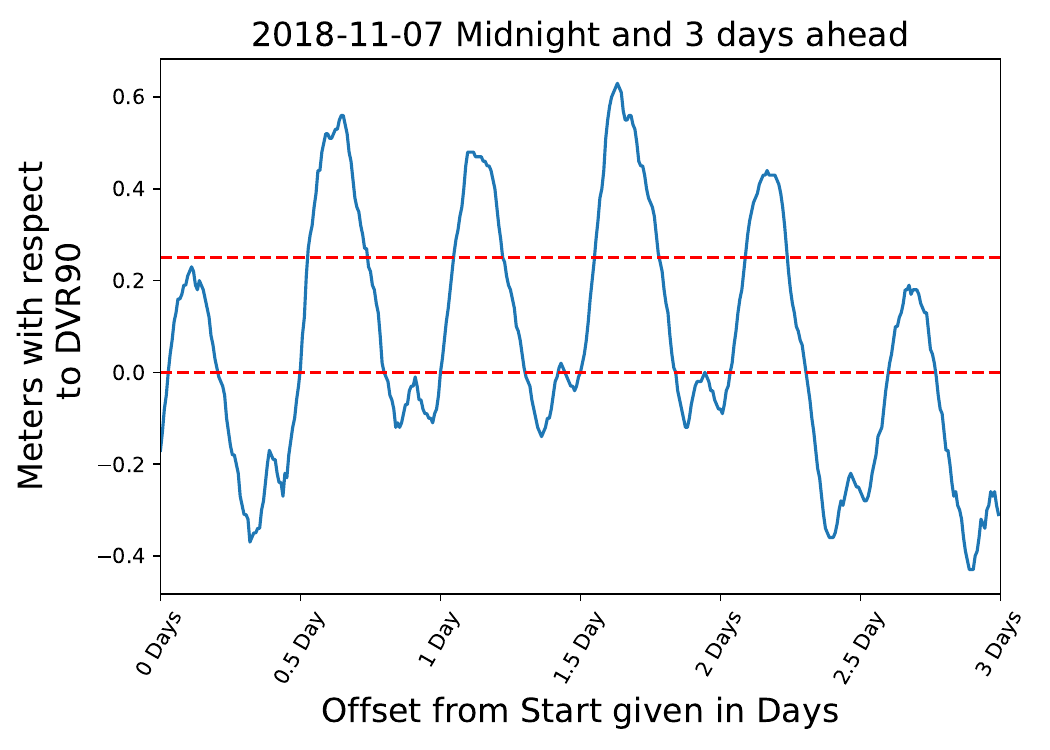}
	\caption{Overall stability data from November.}
	\label{fig:normalPeriod}
  \end{subfigure}\\
  \begin{subfigure}[!b]{.48\textwidth}
    \centering
    \includegraphics[width=\textwidth,height=65mm,keepaspectratio]{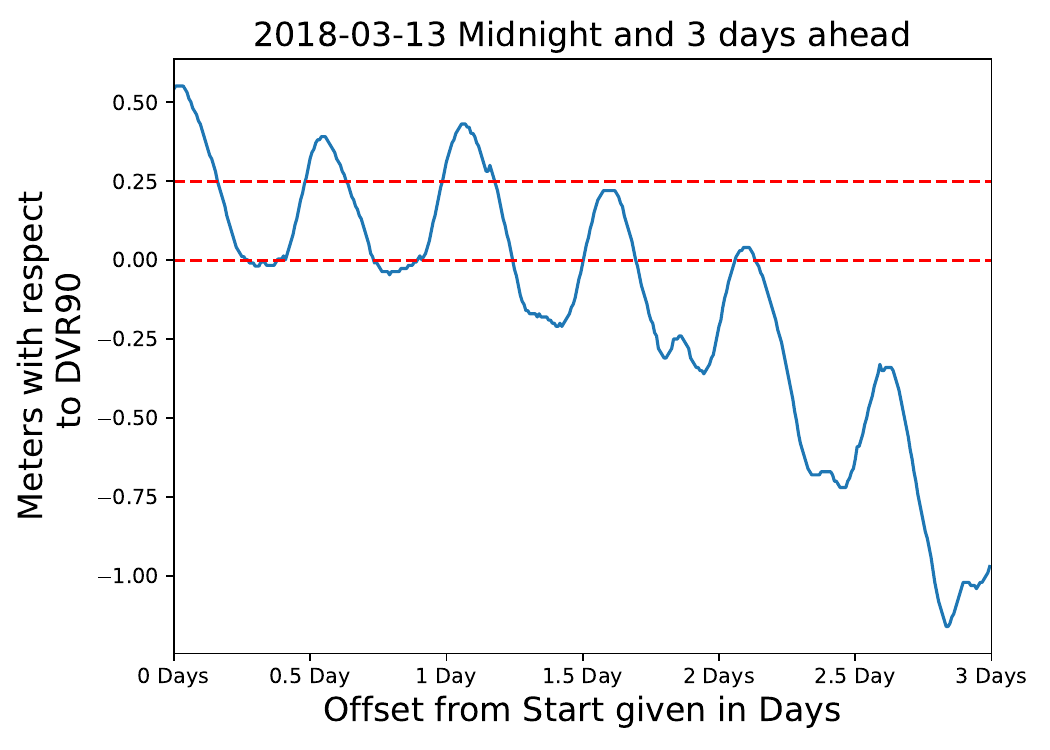}
    \caption{Sharp fall; data from March.}
    \label{fig:lowPeriod}
  \end{subfigure}\hfill%
  \begin{subfigure}[!b]{.48\textwidth}
    \centering
    \includegraphics[width=\textwidth,height=65mm,keepaspectratio]{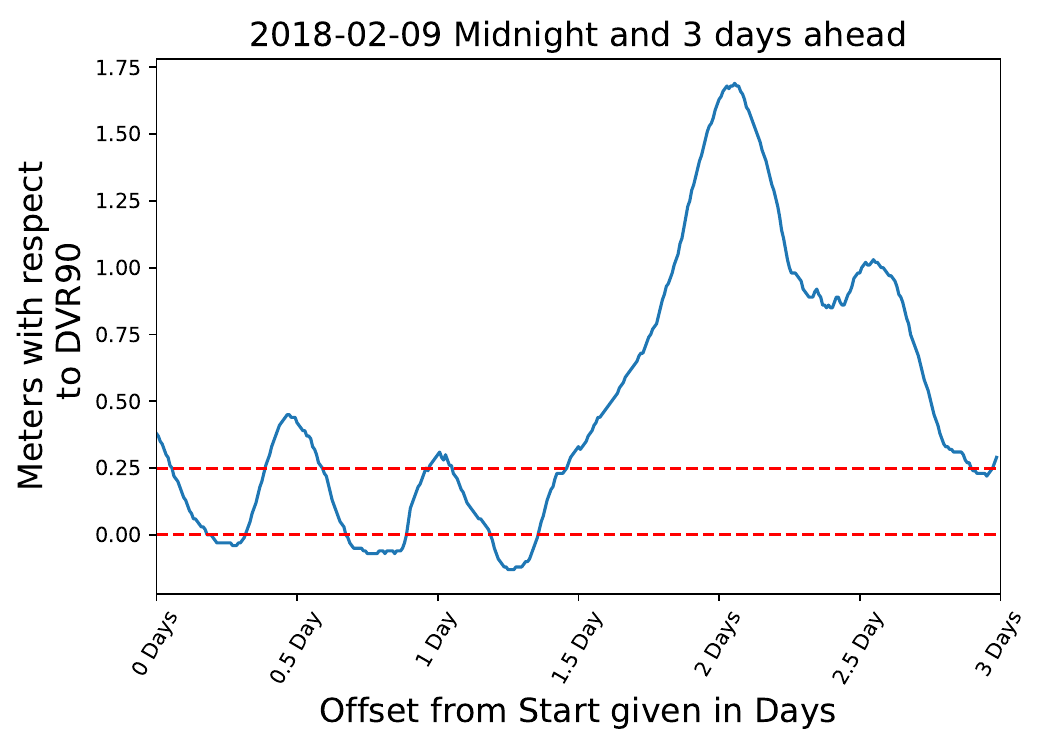}
    \caption{Sharp rise; data from February.}
    \label{fig:highPeriod}
  \end{subfigure}
  \caption{Three-day profiles for the sea levels.}
\end{figure}

\section{Experimental Setup}

In this section, we present how uncertainty of the weather forecast is implemented in the modeling of the system.
Moreover, we run experiments on three different sea-level scenarios, each representing distinct behaviors.
Finally, we present a baseline controller used for comparison.

\subsection{Uncertainty in Forecasts}

Our evaluation is based on historical data.
Instead of using historical weather forecasts, we use the actual historical data of the sea's water level and of the wind speed as if it was the forecast, but we add an uncertainty to it as follows: whenever we update the sea level and the wind in the environment simulation via the \uppFunc{updateSeaAndWind()} function (cf.\ \cref{fig:learner}), we pick a value sampled from a uniform distribution around the actual data.
For the sea's water level, the uncertainty interval is $[-0.005, 0.005]$, while for the wind speed it is $[-0.5, 0.5]$.
Note that we do not add any uncertainty to the historical data when evaluating the controller after learning.

\subsection{Scenarios With Different Sea Levels}

For the learning and evaluation of a controller, we use three 3-day periods from the historical data in 2018, each providing particular scenarios (or profiles) for the sea's behavior.
Each scenario starts at midnight on its given date.
The first period is from November as shown in \cref{fig:normalPeriod}, where we see the typical tidal behavior of the sea level without major outliers.
The tide cycle takes a little more than twelve hours.
In a scenario like this, a controller has plenty of opportunity to add or remove water from the fjord.
As part of the figure, two lines are added to show the band between 0.0 and \qty{0.25}{DVR90}, which is the range where the fjord's water level is deemed safe.
In this range, the controller is able to drain water and, if the wind speed allows it, let water into the fjord, as discussed previously (cf.\ \cref{tab:requirements}).

The two other profiles start with a similar pattern as the first profile, but then either experience low (\cref{fig:lowPeriod}) or high (\cref{fig:highPeriod}) sea levels.
In those scenarios, the controller has some time to adjust the fjord's water level before the sea's water level inhibits the gates from opening.
Recall that, even if the gates are closed, there is still inflow from the surrounding streams.
Both profiles show more than 24 hours where the sea's level is continuously above or below the fjord's threshold levels.

\begin{figure}[ht!]
	\centering
	\includegraphics[width=\textwidth]{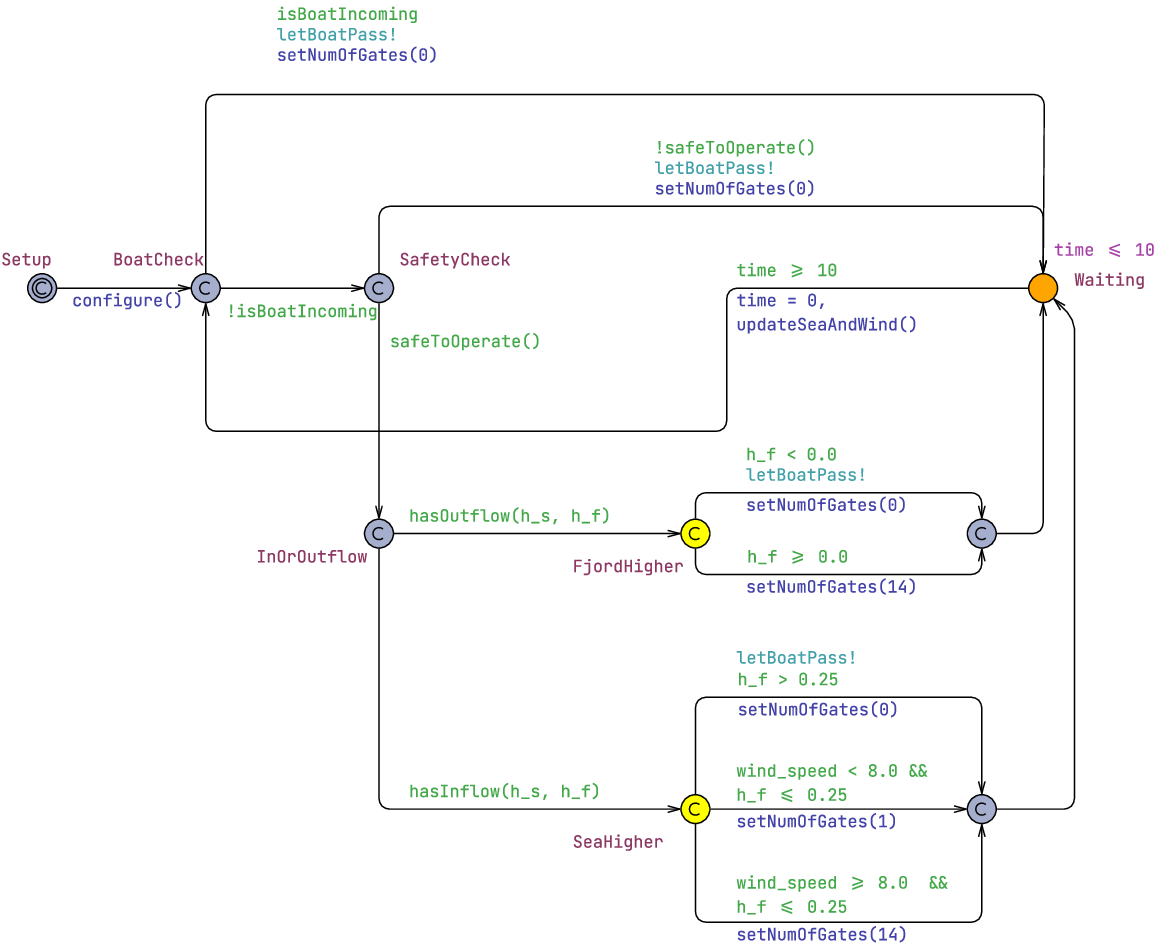}
	\caption{Automaton representing the behavior of the baseline controller. This controller is deterministic, i.e., exactly one transition is enabled in the yellow locations.}
	\label{fig:baseline}
\end{figure}

\subsection{Baseline Controller}

To compare the effectiveness of the learned controller, we use a baseline controller.
The real system is managed by several human operators over time, and there is no formally defined algorithm for controlling the gates, and each operator might control the gates a little differently.
To mitigate the variation of human operators, for the experiments in this work, we have (in collaboration with the DCA) defined a baseline controller that follows the internal guidelines.
The baseline controller is deterministic.
For better comparison with the learned controller, it uses the same control period of ten minutes between choosing an action and receives the same inputs (i.e., the current water levels in the fjord and sea as well as the current wind speed).
Thus, the weather forecast is not considered in the baseline controller.

A \uppaal \stratego automaton implementing the baseline controller's behavior is shown in \cref{fig:baseline}, which we informally describe next.
First, it checks if there are any incoming boats, in which case all gates are closed; this caters to performance requirement~P.3.
If there is no boat, the difference in the water levels must be less than \qty{1}{m} to allow operation according to safety requirement~S.3.
If that is the case, the chosen action depends on the direction of the flow.
If the flow is to the sea, all gates are opened if the fjord's water level is above \qty{0.0}{DCVR90}, and closed otherwise.
If the flow is to the fjord, the gate configuration depends on the wind speed.
If the fjord's level is below \qty{0.25}{DVR90}, then all gates are opened if the wind speed is greater or equal to \qty{8}{m/s}, and otherwise a single gate is opened to allow for fish migration, catering to performance requirement~P.1.
If the fjord's level is above \qty{0.25}{DVR90}, all gates are closed.
This way, safety requirement~S.1 concerning safe water levels is respected.

\subsection{Learning Parameters and Weights}

Before executing the learning query, we must determine a few parameters, such as the weights used in that query.
For the safety range, we choose the weight $w_1 = 10^6$.
This weight is much higher than all other weights because it corresponds to a safety requirement (whereas the three other terms in the cost function only correspond to performance requirements).
The weight for the accumulated waiting time for boats is $w_2 = 0.1$ to ensure that the exponential growth of the cost does not rise too sharply.
The weight for the number of times the gate configuration changes is $w_3 = 20$, assigning a significant penalty to discourage frequent changes to the gate configuration.
The final weight corresponding to the absence of fish migration (despite being possible under the hard constraints) is $w_4 = 1$.
Thus, there is a penalty for not opening any gates for fish, but frequent opening and closing is worse since the weight for changing the gate configuration is greater.

Finally, in \uppaal \stratego, one can adjust the number of learning episodes, along with other parameters~\cite{onTime}.
In our case, we learn three independent controllers, using 6000 episodes each time, and then choose the best-performing controller.
Learning multiple controllers is an effective way to deal with occasional cases where the learning algorithm gets stuck in a local optimum.

\section{Experimental Results}

For each 3-day scenario, we learn a controller via the query presented in \cref{sec:modeling}.
The experiments were conducted on a computer with an Intel(R) Core(TM) Ultra 7 265H CPU running Ubuntu 24.04.
Learning a single controller took approximately 60 minutes.
A reproducibility package is available online~\cite{artifact}.

We then evaluate both the learned controller and the baseline controller in \uppaal \stratego by simulating the same 3-day scenario under each of these controllers.
The properties we measure are the percentage of time within the safe water-level range, the percentage of time where fish were allowed to migrate (provided that the gates were allowed to be opened by the hard constraints), the maximum waiting time that boats had before sailing towards the sluice lock, and the number of gate operations.
Although the sea level and the wind speed are deterministic in the evaluation (since we use the historical data), the arrival of boats is still stochastic because we do not have historical data available.
Thus, we conduct 100 simulations in each setting to obtain the averaged results shown in \cref{tab:results}.

\begin{table}[ht!]
	\caption{Experiment results of the baseline controller and the learned controllers in three scenarios.}
	\label{tab:results}
	\centering
	\begin{tabular}{r|cccc}
      \toprule
      \textbf{Controller} & \textbf{Safety} & \textbf{Fish Migration} & \textbf{Max Waiting Time (minutes)} & \textbf{Gate Operations}  \\
      \midrule
      \multicolumn{5}{l}{\textbf{\textit{Normal Waters (\cref{fig:normalPeriod})}}}  \\
      Baseline & 63.2\% & 95.70\% & 8.95 $\pm$ 0.18 & 26.15 $\pm$ 0.92  \\
      Learned & 100\% & 99.05\% & 8.87 $\pm$ 0.23 & 40.30 $\pm$ 0.00 \\
      \midrule
      \multicolumn{5}{l}{\textbf{\textit{Low Waters (\cref{fig:lowPeriod})}}}  \\
      Baseline &  82.03\%  & 99.62\%  & 9.00 $\pm$ 0.24 & 36.14 $\pm$ 0.93  \\
      Learned & 100\%  & 99.50\%  & 8.83 $\pm$ 0.23 & 37.36 $\pm$ 1.17  \\
      \midrule
      \multicolumn{5}{l}{\textbf{\textit{High Waters (\cref{fig:highPeriod})}}}  \\
      Baseline & 76.9\% & 95.62\%  & 8.89 $\pm$ 0.24    & 26.42 $\pm$ 0.91   \\
      Learned &  100\%  & 96.41\%  & 8.81 $\pm$ 0.24    & 32.43 $\pm$ 0.75   \\
      \bottomrule
	\end{tabular}
\end{table}

\begin{table}[th!]
	\caption{Minimum and maximum fjord levels for the baseline controller and the learned controllers in three scenarios. The numbers are given in meters with respect to DVR90. Violations of the safety range (\qty{0.0}{m}--\qty{0.25}{m}) are marked in bold face.}\label{tab:MinMaxLevels}
	\centering
	\begin{tabular}{c|c|cc}
      \toprule
      \textbf{Sea Level Scenario} & \textbf{Controller} & \textbf{Estimated Min Level} & \textbf{Estimated Max Level} \\
      \midrule
      \multirow{2}{*}{Normal Waters (\cref{fig:normalPeriod})} &
      Baseline & \textbf{$\mathbf{-0.001}$ DVR90} & \qty{0.073}{DVR90}  \\
      & Learned & \qty{0.040}{DVR90} & \qty{0.076}{DVR90} \\
      \midrule
      \multirow{2}{*}{Low Waters (\cref{fig:lowPeriod})} &
      Baseline &  \textbf{$\mathbf{-0.002}$ DVR90}  & \qty{0.147}{DVR90}   \\
      & Learned & \qty{0.035}{DVR90}  & \qty{0.147}{DVR90}   \\
      \midrule
      \multirow{2}{*}{High Waters (\cref{fig:highPeriod})}  &
      Baseline & \qty{0.055}{DVR90} & \textbf{$\mathbf{0.270}$ DVR90}  \\
      & Learned &  \qty{0.060}{DVR90}  & \qty{0.172}{DVR90}   \\
      \bottomrule
	\end{tabular}
\end{table}

First, we observe that the learned controllers are safe all the time.
In contrast, in all three scenarios, the baseline controller violates the safety requirements for multiple hours in total.
In particular, in the first (and most common) scenario, the baseline controller leads to a fjord level outside the safety range more than one third of the time.
In terms of fish migration and the maximum waiting time, the controllers show very similar performance.
They both prioritize quick passage for boats and allow for fish migration most of the time; notably, the learned controller allows for significantly longer migration in the first scenario.
For the number of operations, the baseline controller conducts fewer changes than the learned controllers, in particular in the first scenario (ca.\ 50\% more gate operations) and in the third scenario.

We present the minimum and maximum water levels during the simulations in \cref{tab:MinMaxLevels}.
The baseline controller breaches the safety range by at most \qty{2}{cm}.
As seen before, the learned controllers always stay inside the safety range.
The closest any of the learned controllers are to the minimum or maximum bounds is \qty{3.5}{cm} (during the low-water scenario).

However, the results represent just a single controller for each scenario.
Therefore, they do not give any information on the stability of learning a good controller or whether the chosen weights are optimal.
For instance, the weight for the safety requirement is defined to be much greater than all other weights, as this is the must important property to satisfy.
A drawback of choosing a large weight may be that the learning needs more time to distinguish between the choices that have a relatively small penalty.
In this case, the big penalty of being outside the safe range could make the learner neglect the cost of operating the gate, say, 30 times versus 60 times.
Thus, it is interesting to investigate how much we can lower the weight for the safety requirement~S.1 and still have a safe water level.

\subsection{Weight Tuning}

\begin{figure}[ht!]
	\centering
	\begin{subfigure}[t]{.48\textwidth}
		\centering
		\includegraphics[width=\textwidth,height=65mm,keepaspectratio]{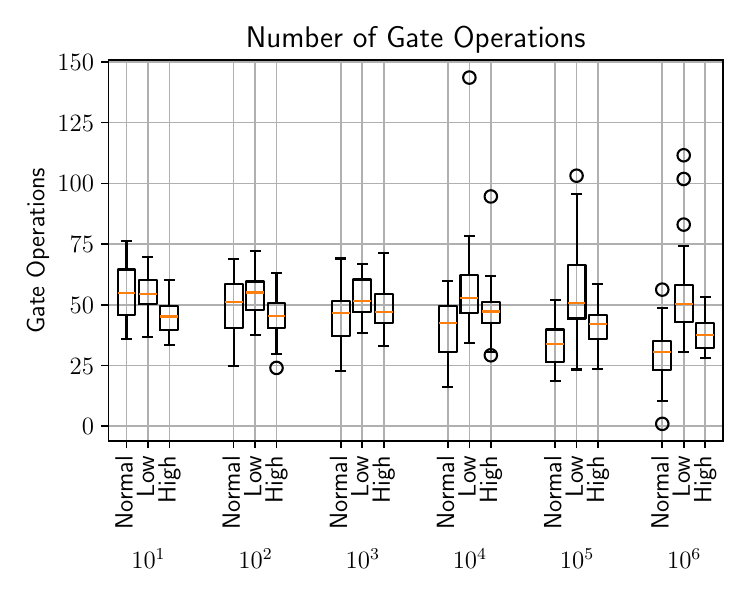}
		\caption{Expected number of gate operations.}
	\label{fig:gates}
	\end{subfigure}\hfill%
	\begin{subfigure}[t]{.48\textwidth}
		\centering
		\includegraphics[width=\textwidth,height=65mm,keepaspectratio]{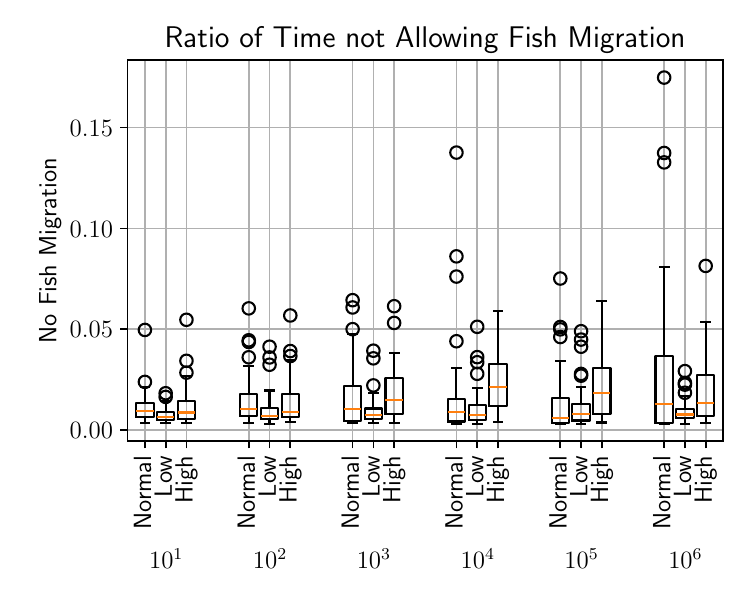}
		\caption{Expected ratio of the time where gates are not opened for fish migration when conditions allow it.}
		\label{fig:migration}
	\end{subfigure}\\
	\begin{subfigure}[t]{.48\textwidth}
		\centering
		\includegraphics[width=\textwidth,height=65mm,keepaspectratio]{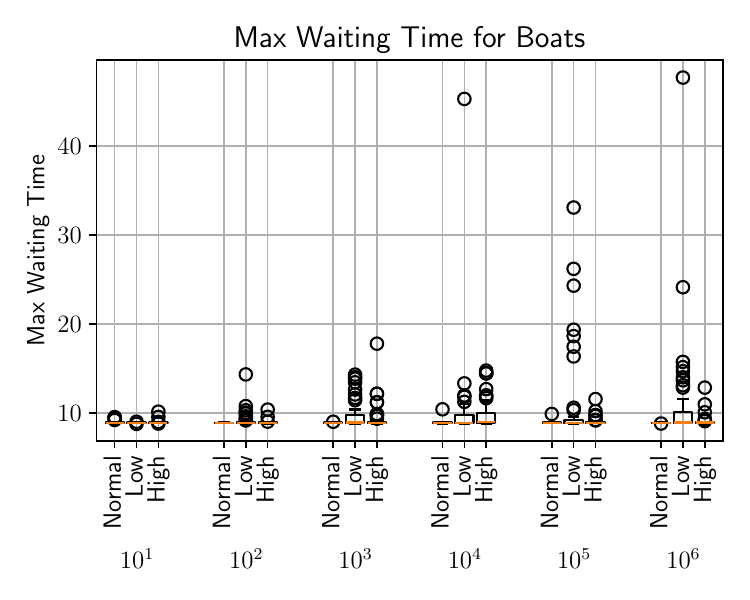}
		\caption{Expected maximum waiting time to close gates for safe passage of boats.}
		\label{fig:waitingTime}
	\end{subfigure}\hfill%
	\begin{subfigure}[t]{.48\textwidth}
		\centering
		\includegraphics[width=\textwidth,height=65mm,keepaspectratio]{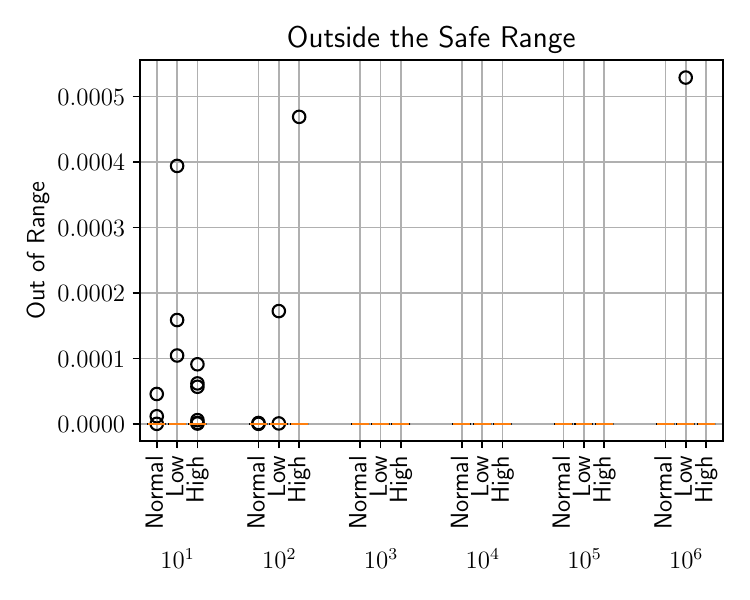}
		\caption{Expected ratio of time outside a safe water level.}
		\label{fig:safe}
	\end{subfigure}
	\caption{Boxplots of the expected performance of controllers generated under different water-level scenarios \textit{(normal, low, high)}. The safety weight~$w_1$ of the learning query ranges from~$10^1$ to~$10^6$.}
	\label{fig:safetyCost}
\end{figure}

To investigate a good value for the safety weight~$w_1$ in our learning query, we run a new experiment where we vary the value of~$w_1$ but keep the values of~$w_2$--$w_4$ fixed (namely the values used to learn the controllers presented in \cref{tab:results}).
The weight~$w_1$ ranges from~$10^1$ to~$10^6$, increasing by one order of magnitude at each step, yielding six configurations.
For each configuration, we learn 50 controllers for each water-level scenario.
We show the expected performance of all controllers in \cref{fig:gates,fig:migration,fig:waitingTime,fig:safe}.

The expected number of gate operations for the normal water-level scenario is decreasing as the value of~$w_1$ increases.
For the two other scenarios, we see no clear indication of which weight is best.
For fish migration, a higher weight increases the expected time of not opening the gates even though the conditions allow it, especially for the normal and high water-level scenarios.
The maximum waiting time of boats increases as~$w_1$ increases.
When the maximum value is above 10~minutes, the controller has chosen not to close the gates initially, thus not prioritizing the waiting boat.
In the worst cases, a boat must wait for more than 40~minutes, which likely is unacceptable in practice (in which case the value of that weight should also be investigated).
Finally, the worst-performing controller violates the safety requirement by being outside the safe water-level range for ${\approx}\,0.55\text{\textperthousand}$ for the time.
Given that the violation is over a short time frame and in the range of millimeters, it may be acceptable in practice.
In future work, a good value for~$w_1$ could be~$10^3$, as almost all violations of the safety requirement happen when using a lower value.
We leave it to future work to identify optimal values for~$w_2$--$w_4$.
However, the optimal weights depend on what the system's owner considers a good balance among the requirements, as different weights lead to different balances, making the choice of weights a political matter.

\section{Conclusion and Future Work}

In this paper, we have demonstrated how \uppaal \stratego can utilize a digital twin of the water level of Ringk\o{}bing Fjord to learn control strategies for operating the gates.
We learn a strategy in $\sim60$ minutes, which is much shorter than the three days a forecast concerns, or the six hours between updates of the forecast, thus demonstrating feasibility.
In our evaluation, the baseline controller was unable to stay within the safety range for the fjord's water level, but all the learned controllers managed to stay within the range all the time.
The controllers' performance for fish migration and boat waiting time was comparable, but all learned controllers increased the number of operations.
Moreover, we demonstrated how the weight used to satisfy the safety requirement can be tuned.
In conclusion, using \uppaal \stratego in an online fashion is a promising approach to optimizing the operation of the gates.
Nonetheless, several aspects must be considered in future work to have an impact on the real system.

\paragraph{Available actions.}
As discussed, we restricted our controller to three possible gate configurations: no open gate, one open gate, or all gates open.
While this corresponds to the standard operation in practice, the digital twin allows us to explore the potential of more fine-grained control, which may allow us to reduce the amount of gate operations.

\paragraph{Optimal weights.}
One possible direction is to compute a Pareto front of the weights used to express the influence of the different terms in the cost function.
In this work, we provide an intuitive reasoning of why the weights are chosen, along with an analysis of the weight used for safety.

\paragraph{Shielding.}
Another promising direction we aim to investigate in the future is the integration of a shield, which prohibits actions that can lead to an unsafe state.
In previous work, we have demonstrated that this is feasible in similar scenarios~\cite{shielding}.

\paragraph{Salinity.}
In practice, another aspect not explicitly considered in this paper is to ensure that the salinity in the fjord stays within a given range to ensure good wildlife conditions.
This further constrains how much seawater and freshwater from the streams should be allowed in a given period, and moreover, how well the salinity gets mixed by the wind.
In this work, we implicitly manage the salinity of the fjord by only allowing water to enter if the wind speed is above \qty{8}{m/s}.
However, we do not keep track of the salinity in our digital twin.
Thus, an explicit integration of salinity into our model is a natural next step.

\paragraph{Water surface.}
Our model assumes a flat water surface of the fjord.
However, in practice, strong winds can push water to one end of the fjord, which may result in sensor readings that differ by \qty{70}{cm}.
Thus, we aim to extend our model to a more precise estimate of the water level in different areas of the fjord.

\paragraph{Inflow from streams.}
A nice feature of the \uppaal \stratego tool is that it allows loading external libraries, which enables co-simulation~\cite{stompc}.
This would allow us to use a more sophisticated tool that models water inflow from the streams, e.g., as in~\cite{larsen2014results}.

\bibliographystyle{eptcs}
\bibliography{bib}





\end{document}